\documentclass[aps,prl,groupedaddress]{revtex4}
\usepackage{graphicx}
\usepackage{amsmath,amssymb,amsthm}
\DeclareMathOperator{\grad}{grad}

\begin{document}

\title{Measurement of bitumen viscosity in the room-temperature drop experiment: student education, public outreach and modern science in one}
\author{A. T. Widdicombe$^{1}$}
\author{P. Ravindrarajah$^{1}$}
\author{A. Sapelkin$^{1}$}
\author{A. E. Phillips$^{1}$}
\author{D. Dunstan$^{1}$}
\author{M. T. Dove$^{1}$}
\author{V. V. Brazhkin$^{2}$}
\author{K. Trachenko$^{1}$}
\address{$^1$ School of Physics and Astronomy, Queen Mary University of London, Mile End Road, London, E1 4NS, UK, email k.trachenko@qmul.ac.uk}
\address{$^2$ Institute for High Pressure Physics, RAS, 142190, Moscow, Russia}

\begin{abstract}
Slow flow of the viscous liquid is a thought-provoking experiment that challenges students, academics and public to think about some fundamental questions in modern science. In the Queensland demonstration, the world-longest running experiment earning the Ig Nobel prize, one drop of pitch takes about 10 years to fall, leading to problems of actually observing the drops. Here, we describe our recent demonstration of slowly-flowing bitumen where appreciable flow is observed on the time scale of months. The experiment is free from dissipative heating effects and has the potential to improve the accuracy of measurement. Bitumen viscosity was calculated by undergraduate students during the summer project. The worldwide access to the running experiment is provided by webcams uploading the images to a dedicated website, enhancing student education experience and promotion of science. This demonstration serves as an attractive student education exercise and stimulates the discussion of fundamental concepts and hotly debated ideas in modern physics research: difference between solids and liquids, the nature of liquid-glass transition, emergence of long time scales in a physical process, and the conflict between human intuition and physical reality.
\end{abstract}

\maketitle

\section{Introduction}

Viscosity is an important property of matter measuring its ability to flow under external perturbation. Most commonly discussed in liquids, viscosity has been studied in other states of matter too: in gases well over the century ago \cite{max}, in solids where flow is often referred to as ``creep''  \cite{frenkel} and, more recently, in exotic states of matter such as quark-gluon plasma \cite{quark}.

Viscosity of liquids is particularly interesting from the physical point of view because liquids are the least understood state of matter in comparison to gases and solids. Surprising though it may seem in this age of scientific advancement, we do not understand even the most basic and fundamental properties of liquids such as their specific heat. Landau and Lifhistz state twice in their statistical physics textbook that liquid heat capacity can not be calculated in general form \cite{landau}, contrary to solids and gases, for the reason that interactions in a liquid are strong and system-specific so that the energy and heat capacity apparently become system-specific too. Strong interactions are successfully treated in solids in the phonon approach, but this has been thought to be inapplicable to liquids where atomic displacements are large.

Liquids, strongly-interacting disordered systems with dynamics intermediate between gases and solids \cite{phystoday}, continue to intrigue not only researchers but also lecturers. In an amusing story about his teaching experience, Granato recalls living in fear about a potential student question about liquid heat capacity \cite{granato}. Observing that the question was never asked by a total of 10000 students, Granato proposes that ``...an important deficiency in our standard teaching method is a failure to mention sufficiently the unsolved problems in physics. Indeed, there is nothing said about liquids [heat capacity] in the standard introductory textbooks, and little or nothing in advanced texts as well. In fact, there is little general awareness even of what the basic experimental facts to be explained are.''

In liquids, viscosity $\eta$ is directly related to the liquid relaxation time $\tau$, the average time between two consecutive atomic jumps at one point in space in the liquid \cite{frenkel}: $\eta=\tau G_{\infty}$, where $G_{\infty}$ is the infinite-frequency shear modulus. Relaxation time provides a simple picture of microscopic dynamics that governs the properties of viscous flow. In liquids, measurable $\tau$ ($\eta$) changes with temperature in a very wide range, spanning 16 orders of magnitude: from $\tau\approx 0.1$ ps as in low-viscous liquids such as room-temperature water or mercury to about 10$^3$ s at which point the liquid stops flowing at the experimental time scale and forms the glass by definition (if crystallization is avoided) \cite{dyre}.

Different types of experiments can be devised to measure $\eta$ and $\tau$. Perhaps surprisingly, the original falling ball Stokes experiment remains popular and able to uncover some non-trivial effects in liquids related to phase transformations \cite{bra}. An alternative way to directly observe viscous flow is to look at the motion of the liquid itself under gravity.

The experimental demonstration of slow pitch flow under gravity has been used for educational and science outreach purposes. Two notable examples include the experiment in the University of Queensland in Australia that started in 1927. This demonstration has been hailed the ``world's longest-running experiment'' and was awarded the Ig Nobel prize \cite{nature2}. A similar demonstration started in Trinity College Dublin in 1944 \cite{nature1}. In these experiments, drops of pitch form every approximately every 10 years \cite{nature2,1984,nature1}. These experiments have attracted the interest of major international news agencies, promoting and popularizing science, an exercise that becomes increasingly important.

The main idea of the slow pitch flow demonstration is to confront long-time physical effects with human experience and intuition acquired at shorter times: pitch appears to be resistant to shear stress and immovable at short time scales and can even be shattered with a hammer, yet it flows at longer time scales as a liquid.

Apart from demonstration and outreach purposes, measurements of slow viscous flow have an important application. The ability to measure viscosity of bitumen reliably has been discussed in Ref. \cite{moran}. In order to avoid dissipative heating present in common viscosity measurements, slow deformation of bitumen droplets was carefully measured. The measured viscosity was found to be an order of magnitude higher than in the previous measurements that involved dissipative heating \cite{moran}.

Our rationale for installing a running bitumen experiment in the School of Physics and Astronomy in Queen Mary University of London was based on several ideas. First, stimulated by the media interest in our work on viscous flow and glass transition \cite{metro}, we have set out to make our idea visual, an important way to engage the students and enhance their learning experience as well as carry out a public outreach exercise.

Second, we sought to have a demonstration that challenges most fundamental concepts of our students, namely differences between states of matter and between liquids and solids in particular. The idea that the transition between the two phases can be continuous rather than discontinuous and that the difference between solid glasses and liquids is only quantitative but not qualitative is intriguing, and has important scientific implications. The idea about the continuous nature of transition between solids and liquids was proposed by Frenkel a long time ago \cite{debate1}. The idea has been quickly criticized by Landau \cite{debate2,debate3} on the basis that the liquid-crystal transition involves symmetry changes and therefore can not be continuous according to his theory of phase transitions. However, Frenkel was referring to solids in general that include both crystals and glasses, emphasizing glasses in his later discussion \cite{frenkel}. Interestingly, the debate between Frenkel and Landau \cite{debate1,debate2,debate3} took place right in between the launches of the two pitch flow experiments in Queensland and Dublin, suggesting that these questions were of increasing interest to both experimental and theoretical community at the time.

Interestingly, the debate between Frenkel and Landau is still at the centre of the famous glass transition problem. According to the large set of experimental data, liquids and glasses are structurally identical \cite{dyre}, and liquid-glass transition does not involve symmetry changes. Yet heat capacity at the liquid-glass transition changes with a jump, still suggesting a phase transition, possibly to a mysterious unknown glass phase \cite{dyre}. This and related issues currently remain at the forefront of the glass transition problem \cite{dyre}, named as the deepest and most interesting unsolved theoretical problem \cite{anderson} and that attracts recurring media attention (see, for example, Ref. \cite{nyc,metro}).

One does not need to assume the existence of a mysterious distinct glass phase and a phase transition in order to explain the heat capacity jump at the glass transition temperature $T_g$: heat capacity jump is the inevitable result of liquid becoming too viscous to flow during the time of experiment \cite{prb}. When the flow stops at the experimental time scale, a system's thermal and elastic properties change as a result, triggering the jump of heat capacity. In this picture, an equilibrium liquid above $T_g$ becomes a non-equilibrium non-flowing liquid (solid glass) below $T_g$ \cite{prb}.

If the glass is just an extremely viscous liquid, can this non-intuitive concept be illustrated in a visual experiment? An obvious demonstration of this would be a very viscous and slowly flowing liquid that appears as a solid during short time scales (e.g. when an experimenter pushes and deforms the liquid with a pin), but starts flowing as a liquid at longer time scales of weeks, months and so on.

Third and importantly, we intended to observe and follow the bitumen flow at reasonably short time scales. Indeed, in the two pitch drop experiments discussed above, drops of pitch form every approximately every 10 years, and have been actually observed only recently when cameras recording the flow were installed and became operational \cite{nature2,nature1}. Our intention was to demonstrate noticeable bitumen flow (e.g. several grams) during the time scale of one academic year. This way, new students starting in September could observe the flow at the end of the final term in May next year. This applies to short-term visitors as well as associated students studying in the University for one year only.

\section{Experiment and results}

We now describe the design and installation of our experiment. The Total UK company has provided two samples of bitumen of different (unspecified) viscosities (samples 35/50 and 95/25 with reported softening temperatures of $50-58\,^{\circ}{\rm C}$ and $90-100\,^{\circ}{\rm C}$, respectively). We have commissioned five glass tubes that narrow down at one end to thinner tubes (see Figure 1) with different orifice radii $r=2.5, 3, 4, 5$ and $6$ mm. We did not have a reliable indication of the flow time originally, hence our reason for using different orifice radii was to have a wide range of flow times (flow time is proportional to $r^4$, see below), increasing the possibility of observing the flow during a reasonable time window. As an additional way of potentially accelerating the flow, we prepared different weights to fit the glass tubes. The weights were not required in the actual experiment since the observable flow started after approximately one month after the installation. The less viscous bitumen (sample 35/50 with reported softening temperature of $50-58\,^{\circ}{\rm C}$) was melted and kept in the furnace for 12 hours at 90$\,^{\circ}{\rm C}$. The equilibrated bitumen melts were poured into the glass tubes in similar amounts. Room temperature variations were in the common $20-25\,^{\circ}{\rm C}$ range.

\begin{figure}
\begin{center}
{\scalebox{0.35}{\includegraphics{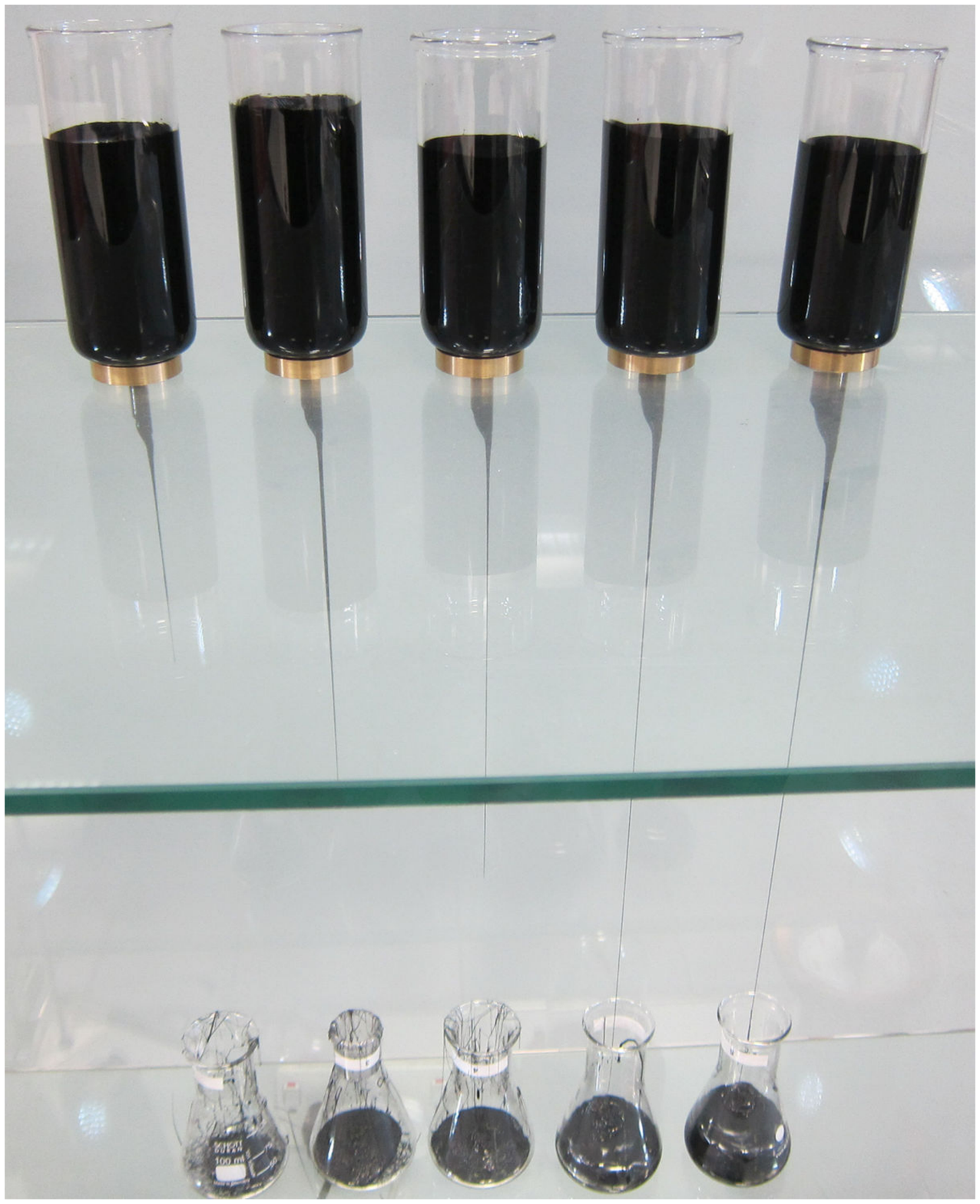}}}
{\scalebox{0.35}{\includegraphics{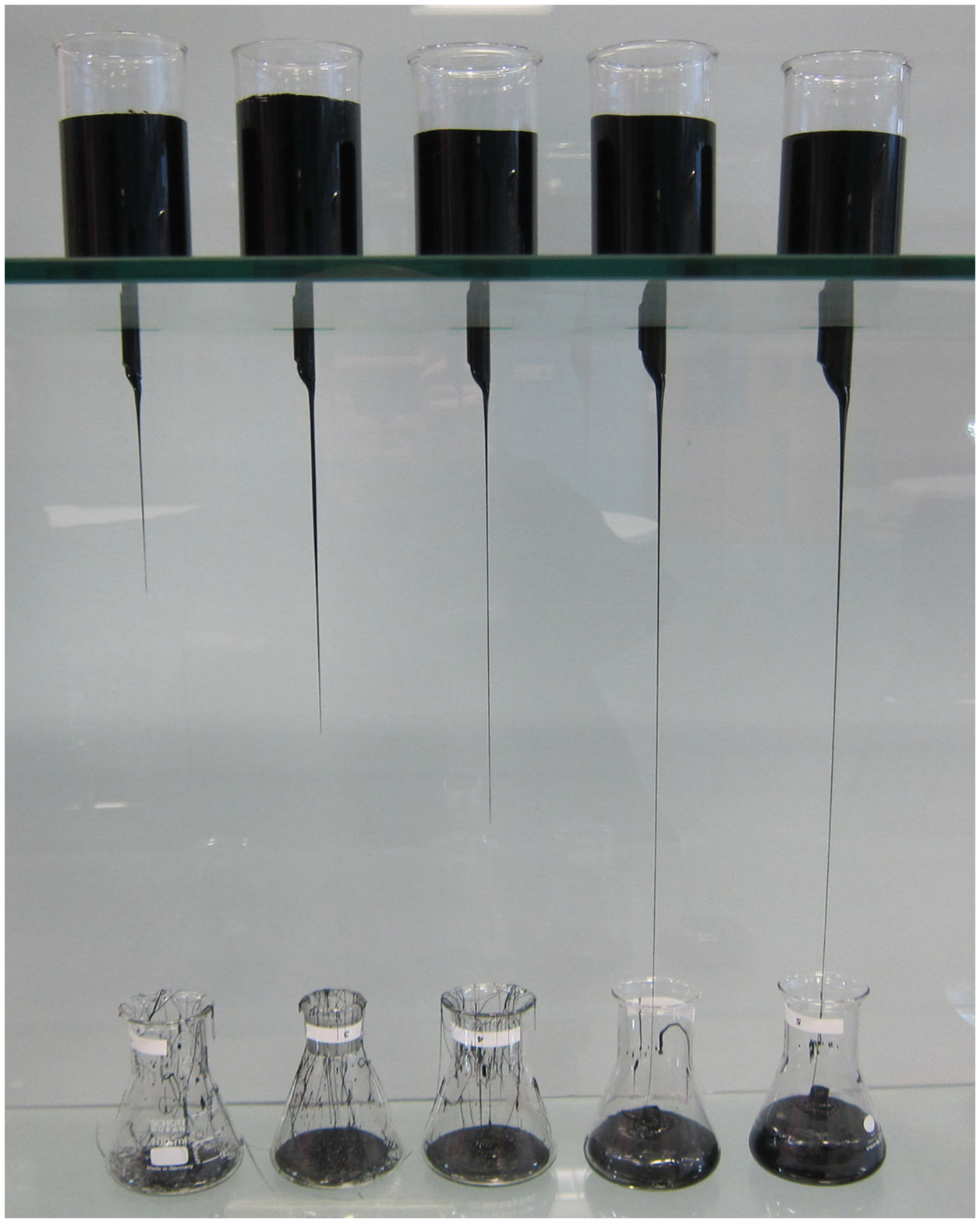}}}
\end{center}
\caption{Experimental setup showing top and bottom views of glass tubes with flowing bitumen. The radius of orifices increases from left to right.}
\label{cv3}
\end{figure}

Two webcams were installed taking pictures every day and downloading them to the file repository connected to the dedicated website \cite{lazy}. A computer programme was written to update and show new photographs every day, copy them into a browsable gallery on the website and to join the photos in the computer animation downloadable from the website. The website \cite{lazy} has served as the additional and important resource of student education as well as science demonstration and outreach. In the digital age, online demonstrations reporting experiments in real time such as this, have the potential to add importantly to the way the education and promotion of science are carried out.

Bitumen viscosity $\eta$ was calculated as part of the student summer project involving our first-year students, the first two authors of this paper. For a tube with radius $r$ and length $L$, $\eta$ and the flow rate are related by the Hagen-Poiseuille equation, the solution of the Navier-Stokes equation for the flow in the circular tube \cite{landau1}:

\begin{equation}
\frac{{\mathrm d}V}{{\mathrm d}t}=\frac{P\pi r^4}{8\eta L}
\label{1}
\end{equation}
\noindent where $V$ is volume, $P$ is the pressure drop across the tube and $\frac{P}{L}$ is the constant pressure gradient.

We consider the flow from the thin bottom sections of the glass tubes with orifice radius $r$ and length $l$ in Figure 1. The flow in the lower thin tube takes place under the hydrostatic pressure $g\rho h$, where $h$ is the height of bitumen in the upper wide section of the tube and $\rho$ is density. In addition, the flow takes place in the gravitational field, giving the extra term $g\rho$ added to the pressure gradient term ($\grad{P}$) in the Euler and Navier-Stokes equations \cite{landau1}. Consequently, the term representing the pressure gradient in the Hagen-Poiseuille equation, $\frac{P}{L}$, modifies as $\frac{g\rho h}{l}+g\rho$:

\begin{equation}
\frac{{\mathrm d}V}{{\mathrm d}t}=\frac{\pi r^4g\rho (l+h)}{8\eta l}
\label{11}
\end{equation}
\noindent

In the flowing bitumen experiment, Eq. (\ref{11}) applies when $P$ or $h$ can be considered approximately constant. This is not the case in our experiment where a considerable part of bitumen has flown out of the tubes (see Figure 1). We therefore write $\frac{{\mathrm d}V}{{\mathrm d}t}$ as $\pi R^2\frac{{\mathrm d}h}{{\mathrm d}t}$, where $R$ is the radius of the upper wide section of the tube. This gives the first-order differential equation, integrating which gives:

\begin{equation}
\eta=\frac{g\rho r^4t}{8R^2l\ln\frac{h_1+l}{h_2+l}}
\label{2}
\end{equation}
\noindent where $h_1$ and $h_2$ are the heights of bitumen in the upper wide part of the glass tube at the beginning and the end of the measurement and $t$ is time of the experiment.

Viscosity $\eta$ was calculated on the basis of data collected in the flow experiment over 317 days. During this time period, the mass of bitumen that flowed and landed in the bottom flasks ranged from 5 g to 53 g for different tubes. This provided enough flow to determine $h_2$ and $h_1$ with reasonable accuracy. We have used Eq. (\ref{2}) to calculate $\eta$ in all five glass tubes. The averaged viscosity was found to be $7.5\times 10^6$ Pa$\cdot$s, with the relative error of 33\%.

Our bitumen sample is therefore about 30 times less viscous than in the world-longest Queensland experiment ($\eta=2.3\times 10^8$ Pa$\cdot$s \cite{nature2,1984}). We note that our and Queensland bitumens are 10--11 orders of magnitude more viscous than room-temperature water. At the same time, they are still 5--6 orders of magnitude less viscous that the liquid at the glass transition where $\eta\approx 10^{13}$ Pa$\cdot$s \cite{dyre}. We comment on this further below.

\section{Discussion and Summary}

We have described an experimental setup that allows students, researchers and public to observe the flow of viscous bitumen during the time period of under one year. This time scale keeps a reasonable balance between fast and slow times: our bitumen appears solid on the time scales of days, yet shows appreciable flow over several months. By providing worldwide access to the experiment and transmitting it in real time, current information technology has the potential to enhance student education experience and public engagement. In view of the current interest in measuring viscosity in slow experiments without dissipative heating \cite{moran}, the proposed experimental setup may be used in further bitumen research and applications.

We make three comments regarding the implications of our demonstration for student education and research. First, the bitumen we used is less viscous than in the previous demonstrations: in the Queensland and Dublin experiments, it takes about 10 years for one drop to fall \cite{nature2,nature1,1984}. Public and media attention to these intriguing and important experiments were drawn primarily because this appears to be unusually long. It is important to remember that ``long'' here refers to human time scales only. From the physical point of view, human time scales do not occupy a special place compared to others. Physical processes in Nature can involve vastly different time scales, and liquid viscosity is a simple demonstration of this: if we cool the familiar SiO$_2$ melt to room temperature (at which it would exist as the glass), liquid relaxation time is calculated to be $10^{67}$ s, 50 orders of magnitude longer than the Age of the Universe \cite{prb}. It is the appreciation of these potential time scales where the additional usefulness of our demonstration lies.

Second, our experiment highlights and contrasts human experience and intuition on one hand and physical laws and processes on the other. A conflict between human experience/intuition and physical laws exists in several areas of physics. Quantum mechanics and general relativity are two notable examples where human experience is in striking conflict with physical laws at vastly different energy and length scales. Slowly flowing bitumen effectively highlights the same conflict but in terms of time: our experience of the apparently solid bitumen, acquired at short observation time, contradicts the liquid behavior at longer times.

Third and finally, this demonstration raises an interesting question related to long time scales in physics: if relaxation time (or flow time) become longer than any practical experiment and hence the experiment only probes the solid properties of the system, is it useful to debate whether the system is a slowly-flowing liquid or a solid? This question brings us back to the original debate between Frenkel and Landau \cite{frenkel,debate1,debate2,debate3} and the continuing discussion about the nature of the liquid-glass transition \cite{dyre,anderson,prb,nyc}. Our answer to this question is positively ``yes'', because the answer has important consequences for the theories and models that we develop to describe the physical world. Namely, the experimental absence of structural and symmetry changes between the liquid and the glass \cite{dyre} imply that previously developed theories and models based on phase transition concepts do not apply. This stimulates the development of new theories and paradigms and therefore drives the field forward. Regardless of what glass transition theories and models become accepted in the future, the simple and accessible demonstration of slow bitumen flow continues to challenge our students and ourselves.

We are grateful to the School of Physics and Astronomy and EPSRC for support, Total UK for providing bitumen samples, G. Simpson for experimental assistance and B. Still, A. Owen, D. Bolmatov P. Micakovic and T. Arter for their support with IT and website.

\end{document}